\documentclass[pra,aps,superscriptaddress,twocolumn,color]{revtex4-1}

\usepackage{amsmath}
\usepackage{amsfonts}
\usepackage{bm}
\usepackage{graphicx}
\usepackage{array,color}

\begin{document}
\title{Faraday patterns generated by Rabi oscillation in a binary
 Bose-Einstein condensate}

%\author{Torigen Chin}
\author{Terun Chen}
\affiliation{Department of Engineering Science, University of
Electro-Communications, Tokyo 182-8585, Japan}

\author{Kosuke Shibata}
\affiliation{Department of Physics, Gakushuin University, Tokyo 171-8588,
  Japan}

\author{Yujiro Eto}
\affiliation{National Institute of Advanced Industrial Science and
  Technology (AIST), NMIJ, Ibaraki 305-8568, Japan}

\author{Takuya Hirano}
\affiliation{Department of Physics, Gakushuin University, Tokyo 171-8588,
  Japan}

\author{Hiroki Saito}
\affiliation {Department of Engineering Science, University of
Electro-Communications, Tokyo 182-8585, Japan}

\date{\today}

\begin{abstract}
The interaction between atoms in a two-component Bose-Einstein condensate
(BEC) is effectively modulated by the Rabi oscillation.
This periodic modulation of the effective interaction is shown to generate
Faraday patterns through parametric resonance.
We show that there are multiple resonances arising from the density and spin
waves in a two-component BEC, and investigate the interplay between the
Faraday-pattern formation and the phase separation.
\end{abstract}

\maketitle

\section{Introduction}
\label{s:intro}

Bose-Einstein condensates (BECs) of ultracold alkali atoms are generated
through magnetic and optical means, and are created in one of several
hyperfine states of the electronic ground state.
When the hyperfine state is coherently coupled with another hyperfine state
by microwave or radio-frequency wave, Rabi oscillation occurs between the
two states, resulting in a two-component (binary) BEC~\cite{Matthews,
  Nicklas, Hamner}.
Coherently-coupled two-component BECs have much richer physics than
single-component BECs, and they have been studied in detail
theoretically~\cite{Williams, Park, Kasamatsu, Lee, Sabbatini, Dror,
  Bernier, Fialko, Usui}.

Recently, a coherently-coupled BEC of the hyperfine
states $|F = 1, m_1 \rangle$ and $|F = 1, m_2 \rangle$ of $^{87}{\rm Rb}$
atoms was experimentally investigated~\cite{Shibata}, where $F$ is the
hyperfine spin and $m_{1,2} = 0, \pm 1$ are the magnetic quantum numbers.
Since $F = 1$ is the lowest-energy hyperfine spin, stable Rabi oscillation
over a long duration was realized with precise control of the coupling
field.
In this experiment, excitation patterns were observed in the atomic cloud
after Rabi oscillation with a small Rabi frequency ($\sim 100$ Hz), which
was also confirmed numerically (Figs.~7(d) and 8 in Ref.~\cite{Shibata}).
This excitation of the BEC can be interpreted as follows.
Suppose that all the atoms are in state 1 or 2 at some instant.
The energy density of the interatomic interaction for this state is $g_{11}
n^2 / 2$ or $g_{22} n^2 / 2$, where $n$ is the atomic density and $g_{ij}$
is the interaction coefficient between states $i$ and $j$.
When states 1 and 2 become equally populated in the course of the Rabi
oscillation, the energy density becomes $(g_{11} + 2 g_{12} + g_{22}) n^2 /
8$.
Thus, the interatomic interaction effectively oscillates in time due to the
Rabi oscillation unless $g_{11} = g_{22} = g_{12}$.
Such an effective modulation of the interaction can be used to control the
dynamics of BECs~\cite{Saito07, Eto}.

In a single-component BEC, the periodic modulation of the interaction by,
e.g., a Feshbach resonance, was shown to excite a pattern in the atomic
density through a parametric resonance, which is called the Faraday
pattern~\cite{Staliunas}.
Such Faraday-pattern formation was experimentally observed by periodic
modulation of the transverse confinement in a cigar-shaped
BEC~\cite{Engels}, which effectively modulates the interaction.
The Faraday waves can also be generated in two-component BECs~\cite{Balaz},
dipolar BECs~\cite{Lakomy}, and Fermi-Bose mixtures~\cite{Abdullaev}.

Motivated by the experiment reported in Ref.~\cite{Shibata}, in the present
paper we theoretically investigate Faraday-pattern formation in a
coherently-coupled two-component BEC.
Based on the oscillation of the effective interaction due to the Rabi
oscillation, Faraday patterns are shown to be excited without any modulation
of external parameters.
Since there are two types of elementary excitations in a two-component BEC,
i.e., density and spin waves, multiple wave-number excitations appear in the
Faraday waves.
In the case of $g_{11} = g_{22}$, we find that three wave numbers are
excited, which arises from the resonance of the Rabi oscillation with
density-density, density-spin, and spin-spin waves.
We show using Floquet analysis that the Faraday-wave excitation can coexist
with phase separation due to the immiscibility of the two components.

The reminder of this paper is organized as follows.
Section~\ref{s:formulation} formulates the mean-field description of a
coherently-coupled two-component BEC.
Section~\ref{s:evolution} presents numerical results of the time evolution
of the system.
Section~\ref{s:floquet} performs the Floquet analysis of the Rabi induced
excitations.
Section~\ref{s:resonance} studies the resonance conditions and provides the
mechanism of multiple resonances, and Sec.~\ref{s:conc} concludes the
study.

\section{Coupled Gross-Pitaevskii equations}
\label{s:formulation}

We consider a two-component BEC in uniform space at zero temperature.
In the mean-field approximation, the condensate is described by the
macroscopic wave functions $\psi_1(\bm{r}, t)$ and $\psi_2(\bm{r}, t)$ for
components 1 and 2, respectively.
We normalize the length, time, and atomic density by an arbitrary length
$L$, time $T$, and density $n_0$, where $\hbar T = M L^2$ is satisfied with
$M$ being the mass of an atom.
The coupled Gross-Pitaevskii (GP) equations are then given by
\begin{subequations} \label{GP}
\begin{eqnarray}
i \frac{\partial\psi_1}{\partial t} & = & -\frac{\nabla^2}{2} \psi_1 +
g_{11} |\psi_1|^2 \psi_1 + g_{12} |\psi_2|^2 \psi_1 - i \Omega \psi_2,
\nonumber \\
\\
i \frac{\partial\psi_2}{\partial t} & = & -\frac{\nabla^2}{2} \psi_2 +
g_{22} |\psi_2|^2 \psi_2 + g_{12} |\psi_1|^2 \psi_2 + i \Omega^* \psi_1,
\nonumber \\
\end{eqnarray}
\end{subequations}
where $g_{ij} = 4 \pi a_{ij} L^2 n_0$ is the nondimensional interaction
coefficient and $a_{ij}$ is the scattering length between atoms in states
$i$ and $j$.
The phase of the Rabi coupling is determined by the phase of the coupling
field, and $\Omega$ is taken to be real and positive without loss of
generality.
In the following study, for simplicity, we assume $g_{11} = g_{22} \equiv
g > 0$ and $g_{12} > 0$.
We also assume that the initial state is the uniform state with equal
population, $\psi_1(t=0) = \psi_2(t=0) = 1 / \sqrt{2}$.

In the noninteracting system ($g = g_{12} = 0$), the uniform solution of
Eq.~(\ref{GP}) is given by
\begin{subequations}
\begin{eqnarray}
\psi_1(t) & = & \cos(\Omega t + \pi / 4), \\
\psi_2(t) & = & \sin(\Omega t + \pi / 4).
\end{eqnarray}
\end{subequations}
We note that the oscillation frequency of the densities $|\psi_1|^2$ and
$|\psi_2|^2$ is $2\Omega$.

In the absence of the Rabi coupling ($\Omega = 0$), the miscibility of the
two components is determined by the interaction coefficients $g_{ij}$.
The frequency $\omega$ of the excitation with wave number $k$ for the
state $\psi_1 = \psi_2 = 1 / \sqrt{2}$ has the form~\cite{Pethick},
\begin{eqnarray} \label{om0}
  \omega^2 & = & \frac{\epsilon_k}{2} \left[ 2\epsilon_k + g_{11} + g_{22}
  \pm \sqrt{(g_{11} - g_{22})^2 + 4 g_{12}^2} \right]
  \nonumber \\
  & = & \epsilon_k (\epsilon_k + g \pm g_{12}),
\end{eqnarray}
where $\epsilon_k = k^2 / 2$ is the kinetic energy of a free particle.
The frequency $\omega$ in Eq.~(\ref{om0}) becomes imaginary for $g_{11}
g_{22} = g^2 < g_{12}^2$, which corresponds to the instability of phase
separation.
The most unstable wave number is $k = \sqrt{g_{12} - g}$.

When the Rabi coupling $\Omega$ is much larger than the energy scale of the
interaction $g_{ij}$, the system is effectively described by the dressed
states $\psi_\pm = (\psi_1 \pm \psi_2) / \sqrt{2}$.
The intra- and inter-component interactions for the dressed states are given
by~\cite{Nicklas, Search, Jenkins, Merhasin}
\begin{subequations} \label{dressg}
\begin{eqnarray}
g_{++} = g_{--} & = & \frac{1}{4}(g_{11} + g_{22} + 2 g_{12}) =
\frac{1}{2}(g + g_{12}), \\
g_{+-} & = & \frac{1}{2}(g_{11} + g_{22}) = g.
\end{eqnarray}
\end{subequations}
Using these effective interaction coefficients $g_{++}$, $g_{--}$,
and $g_{+-}$ in Eq.~(\ref{om0}) instead of $g_{11}$, $g_{22}$, and $g_{12}$,
we find
\begin{eqnarray} \label{opposite}
  \omega^2 = \epsilon_k [\epsilon_k + (3 g + g_{12}) / 2], \;
  \epsilon_k [\epsilon_k + (g_{12} - g) / 2].
\end{eqnarray}
The miscibility condition for large $\Omega$ is therefore opposite to that
for $\Omega = 0$, i.e., the uniformly mixed state $\psi_1 = \psi_2 = 1 /
\sqrt{2}$ is unstable and phase separation occurs for $g > g_{12}$.
The most unstable wave number is $k = \sqrt{(g - g_{12}) / 2}$.

\section{Time evolution of the system}
\label{s:evolution}

We investigate the dynamics of the system by solving the GP equation
(\ref{GP}) numerically.
For simplicity, we consider a two-dimensional (2D) system.
In the numerical calculation, the wave function is discretized into a $2048
\times 2048$ mesh, where each mesh has dimensions $dx = dy = 0.2$, and the
size of the entire 2D space is $409.6 \times 409.6$.
The time evolution is obtained using the pseudospectral
method~\cite{recipe}.
The initial state of $\psi_1(\bm{r})$ and $\psi_2(\bm{r})$ is set to be a
constant $1 / \sqrt{2}$ plus a small random number for each mesh.
This small initial noise breaks the exact numerical symmetry and triggers
spatial pattern formation.
Because the pseudospectral method is used, a periodic boundary condition is
imposed.

\begin{figure}[tb]
\includegraphics[width=8cm]{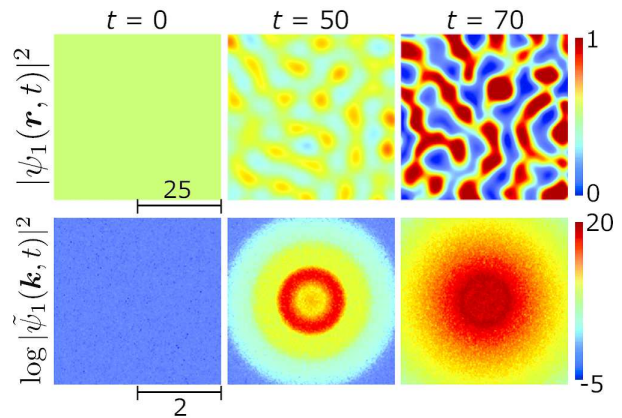}
\caption{
  (color online) Time evolution of the system without Rabi coupling ($\Omega
  = 0$) for the immiscible condition, $g = 1$ and $g_{12} = 1.4$.
  The density profile of component 1 in the real space, $|\psi_1(\bm{r},
  t)|^2$, and that in the momentum space, $\log |\tilde\psi_1(\bm{k},
  t)|^2$, are shown in the upper and lower panels, respectively.
  See the Supplemental Material for a movie illustrating the
  dynamics~\cite{movies}.
}
\label{f:separation}
\end{figure}
First, in Fig.~\ref{f:separation}, we show the case without Rabi coupling
($\Omega = 0$), where the interaction coefficients satisfy the immiscible
condition, $g = 1$ and $g_{12} = 1.4$.
We see that the density pattern emerges in $|\psi_1(\bm{r}, t)|^2$ (and also
in $|\psi_2(\bm{r}, t)|^2$) at $t = 50$, which leads to the phase
separation at $t = 70$.
In this dynamics, the total density $|\psi_1(\bm{r}, t)|^2 + |\psi_2(\bm{r},
t)|^2$ is almost constant with time.
In Fig.~\ref{f:separation}, the logarithmic density profile in the momentum
space, $\log |\tilde\psi_1(\bm{k}, t)|^2$, is also shown, where
\begin{equation}
  \tilde\psi_j(\bm{k}, t) = \sum_{\bm{r}} \psi_j(\bm{r}, t)
  e^{-i \bm{k} \cdot \bm{r}}
\end{equation}
is the discrete Fourier transformation over the numerical mesh.
The ring-shaped peak appears in the momentum distribution due to the
pattern formation.
The radius of the ring at $t = 50$ is $k \simeq 0.4$-$0.8$, which includes
the most unstable wave number predicted from Eq.~(\ref{om0}),
$k = \sqrt{g_{12} - g} \simeq 0.63$.

\begin{figure}[tb]
\includegraphics[width=8cm]{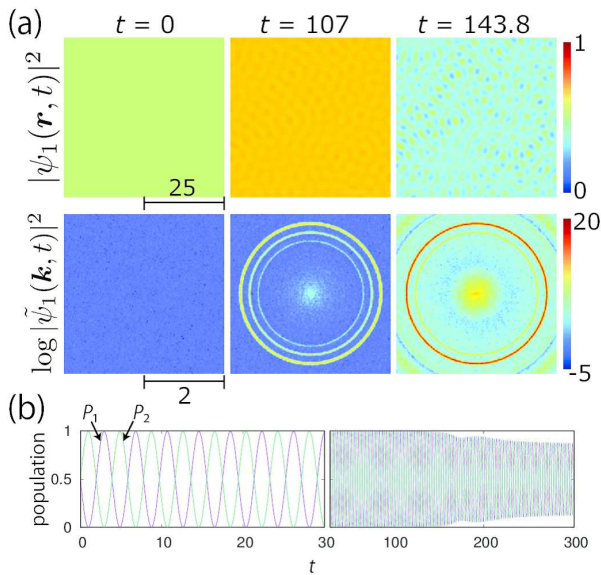}
\caption{
  (color online) Time evolution of the system with Rabi coupling ($\Omega
  = 0.8$) for $g = 1$ and $g_{12} = 1.4$.
  (a) The density profile of component 1 in the real space, $|\psi_1(\bm{r},
  t)|^2$, and that in the momentum space, $\log |\tilde\psi_1(\bm{k}, t)|^2$,
  are shown in the upper and lower panels, respectively.
  See the Supplemental Material for a movie illustrating the
  dynamics~\cite{movies}.
  (b) Time evolution of the population $P_j(t)$ in each component for $0 < t
  < 30$ and $30 < t < 300$.
}
\label{f:three}
\end{figure}
Figure~\ref{f:three} shows the dynamics in the presence of the Rabi coupling
with $\Omega = 0.8$.
Starting from $\psi_1 = \psi_2 = 1 / \sqrt{2}$, the two components undergo
Rabi oscillation with angular frequency $\omega \simeq 1.62$, as shown in
Fig.~\ref{f:three}(b).
Due to the Rabi oscillation, the lifetime of the uniform distribution is
prolonged, compared with that in Fig.~\ref{f:separation} for the same
interaction coefficients $g = 1$ and $g_{12} = 1.4$.
At $t \simeq 100$, the density pattern emerges, as shown in
Fig.~\ref{f:three}(a).
Interestingly, three rings appear in the momentum space at $t \simeq 100$
(i.e., three wave numbers are simultaneously excited) with radii $k \simeq
1.3$, $1.5$, and $1.75$.
Among the three rings, the outermost ring becomes dominant and the innermost
ring disappears at $t \simeq 143$.
In Fig.~\ref{f:three}(b), the time evolution of the population $(j = 1, 2)$
\begin{equation}
P_j(t) = \frac{\int |\psi_j(\bm{r}, t)|^2 d\bm{r}}{\int \left[
    |\psi_1(\bm{r}, t)|^2 + |\psi_2(\bm{r}, t)|^2 \right] d\bm{r}}
\end{equation}
of each component is plotted.
The amplitude of the oscillation in $P_j(t)$ decreases as the pattern
develops, which is due to the inhomogeneous Rabi oscillation.

In the following sections, we will show that the excitation of the three
wave numbers in Fig.~\ref{f:three}(a) is due to the interplay between the
Rabi oscillation and the interatomic interaction.
The population of each component oscillates due to the Rabi oscillation, and
this leads to oscillation of the interaction energy because $g \neq
g_{12}$.
Such oscillation of the effective interaction causes parametric resonance
with Bogoliubov modes, which is reminiscent of Faraday-pattern formation in
a single-component BEC with time-dependent interaction~\cite{Staliunas}.

\begin{figure}[tb]
\includegraphics[width=8cm]{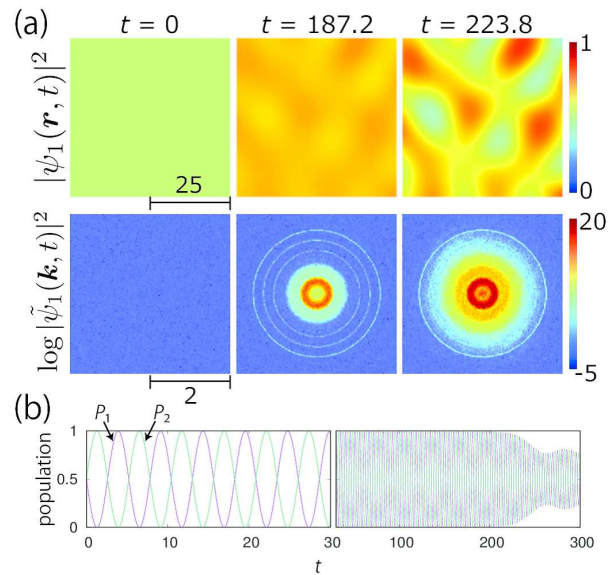}
\caption{
  (color online) Time evolution of the system for $\Omega = 0.6$, $g = 1$,
  and $g_{12} = 0.8$.
  (a) The density profile of component 1 in the real space, $|\psi_1(\bm{r},
  t)|^2$, and that in the momentum space, $\log |\tilde\psi_1(\bm{k}, t)|^2$,
  are shown in the upper and lower panels, respectively.
  See the Supplemental Material for a movie illustrating the
  dynamics~\cite{movies}.
  (b) Time evolution of the population $P_j(t)$ in each component for $0 < t
  < 30$ and $30 < t < 300$.
}
\label{f:miscible}
\end{figure}
Figure~\ref{f:miscible} shows the case of $g = 1$ and $g_{12} = 0.8$, for
which the two components are miscible for $\Omega = 0$ and immiscible for
$\Omega \rightarrow \infty$.
For the intermediate value of $\Omega$, the immiscibility coexists with
the Faraday-pattern formation due to the Rabi oscillation.
At $t \simeq 180$, a thick ring is observed near the center in the momentum
space, in addition to the thin three rings.
The central ring in the momentum space arises from the phase separation of
the dressed states.
According to Eq.~(\ref{opposite}), the most unstable wave number for $\Omega
\rightarrow \infty$ is $k = \sqrt{(g - g_{12}) / 2} \simeq 0.3$, which is in
reasonable agreement with the radius of the central ring.
As shown later, the outer three rings correspond to the Faraday-like
excitation due to the Rabi oscillation.
In Figs.~\ref{f:three} and \ref{f:miscible}, the patterns also appear in
component 2 (data not shown), and the total density $|\psi_1(\bm{r}, t)|^2 +
|\psi_2(\bm{r}, t)|^2$ is almost uniform.

\section{Floquet analysis}
\label{s:floquet}

To study the excitation spectra shown in
Figs.~\ref{f:separation}-\ref{f:miscible} in more detail, we performed
Floquet analysis as follows.
The Rabi oscillation in the uniform system is described by the wave
functions as ($j = 1, 2$)
\begin{equation} \label{unist}
\psi_j(t) = e^{-i \mu t} f_j(t),
\end{equation}
where $\mu$ is a real constant and $f_j(t)$ are periodic complex functions
satisfying $f_j(t+T) = f_j(t)$, with $T$ being the Rabi oscillation period.
We write the small excitation of the state~(\ref{unist}) as
\begin{equation} \label{excite}
\psi_j(\bm{r}, t) = e^{-i \mu t} \left[ f_j(t) + \delta\psi_j(\bm{r}, t)
\right].
\end{equation}
We assume that the excitation has wave number $\bm{k}$,
\begin{equation} \label{dpsi}
  \delta\psi_j(\bm{r}, t) = u_j(t) e^{i \bm{k} \cdot \bm{r}}
  + v_j^*(t) e^{-i \bm{k} \cdot \bm{r}}.
\end{equation}
Substituting Eqs.~(\ref{excite}) and (\ref{dpsi}) into the GP equation
(\ref{GP}), we obtain four coupled differential equations,
\begin{subequations} \label{diffuv}
\begin{eqnarray}
  i \dot{u}_1 & = & \left( \epsilon_k - \mu \right) u_1
  + g \left( 2|f_1|^2 u_1 + f_1^2 v_1 \right) \nonumber \\
  & & + g_{12} \left( |f_2|^2 u_1 + f_1 f_2^* u_2 + f_1 f_2 v_2 \right)
  - i \Omega u_2, \\
  i \dot{v}_1 & = & -\left( \epsilon_k - \mu \right) v_1
  - g \left( 2|f_1|^2 v_1 + f_1^{*2} u_1 \right) \nonumber \\
  & & - g_{12} \left( |f_2|^2 v_1 + f_1^* f_2 v_2 + f_1^* f_2^* u_2 \right)
  - i \Omega v_2, \\
  i \dot{u}_2 & = & \left( \epsilon_k - \mu \right) u_2
  + g \left( 2|f_2|^2 u_2 + f_2^2 v_2 \right) \nonumber \\
  & & + g_{12} \left( |f_1|^2 u_2 + f_1^* f_2 u_1 + f_1 f_2 v_1 \right)
  + i \Omega u_1, \\
  i \dot{v}_2 & = & -\left( \epsilon_k - \mu \right) v_2
  - g \left( 2|f_2|^2 v_2 + f_2^{*2} u_2 \right) \nonumber \\
  & & - g_{12} \left( |f_1|^2 v_2 + f_1 f_2^* v_1 + f_1^* f_2^* u_1 \right)
  + i \Omega v_1.
\end{eqnarray}
\end{subequations}
The coefficients of $u_j(t)$ and $v_j(t)$ on the right-hand side of these
equations are all periodic with period $T$, since $f_1(t)$ and $f_2(t)$
are periodic functions.
Therefore, according to the Floquet theorem, there exist solutions $\bm{u} =
(u_1, v_1, u_2, v_2)^{\rm tr}$ satisfying
\begin{equation} \label{flo}
  \bm{u}(t) = e^{\lambda t} \bm{p}(t),
\end{equation}
where tr indicates transpose, $\lambda$ is the Floquet exponent, and
$\bm{p}(t)$ are periodic functions with period $T$.
At $t = T$, Eq.~(\ref{flo}) becomes
\begin{equation} \label{flo2}
  \bm{u}(T) = e^{\lambda T} \bm{p}(T) = e^{\lambda T} \bm{p}(0).
\end{equation}

We perform the Floquet analysis numerically as follows.
We first obtain the uniform solution~(\ref{unist}) by solving the uniform GP
equation using the Runge-Kutta method, which yields $\mu$, $T$, and
$f_j(t)$.
Next, we integrate Eq.~(\ref{diffuv}) from $t = 0$ to $t = T$ using the
Runge-Kutta method, where the initial values of $\bm{u}$ are $\bm{u}_a(0) =
(1, 0, 0, 0)^{\rm tr}$, $\bm{u}_b(0) = (0, 1, 0, 0)^{\rm tr}$, $\bm{u}_c(0)
= (0, 0, 1, 0)^{\rm tr}$, and $\bm{u}_d(0) = (0, 0, 0, 1)^{\rm tr}$, giving
$\bm{u}_a(T)$, $\bm{u}_b(T)$, $\bm{u}_c(T)$, and $\bm{u}_d(T)$,
respectively.
Using these results, we define a $4 \times 4$ matrix as
\begin{equation} \label{UT}
  U_T = [\bm{u}_a(T), \bm{u}_b(T), \bm{u}_c(T), \bm{u}_d(T)].
\end{equation}
The values of $\bm{u}(t = T)$ for arbitrary initial values $\bm{u}(t = 0)$
can be expressed as
\begin{equation}
\bm{u}(T) = U_T \bm{u}(0).
\end{equation}
The eigenvalues and eigenvectors of $U_T$ thus correspond to $e^{\lambda T}$
and $\bm{p}(0)$ in Eq.~(\ref{flo2}), respectively.

\begin{figure}[tb]
\includegraphics[width=8.5cm]{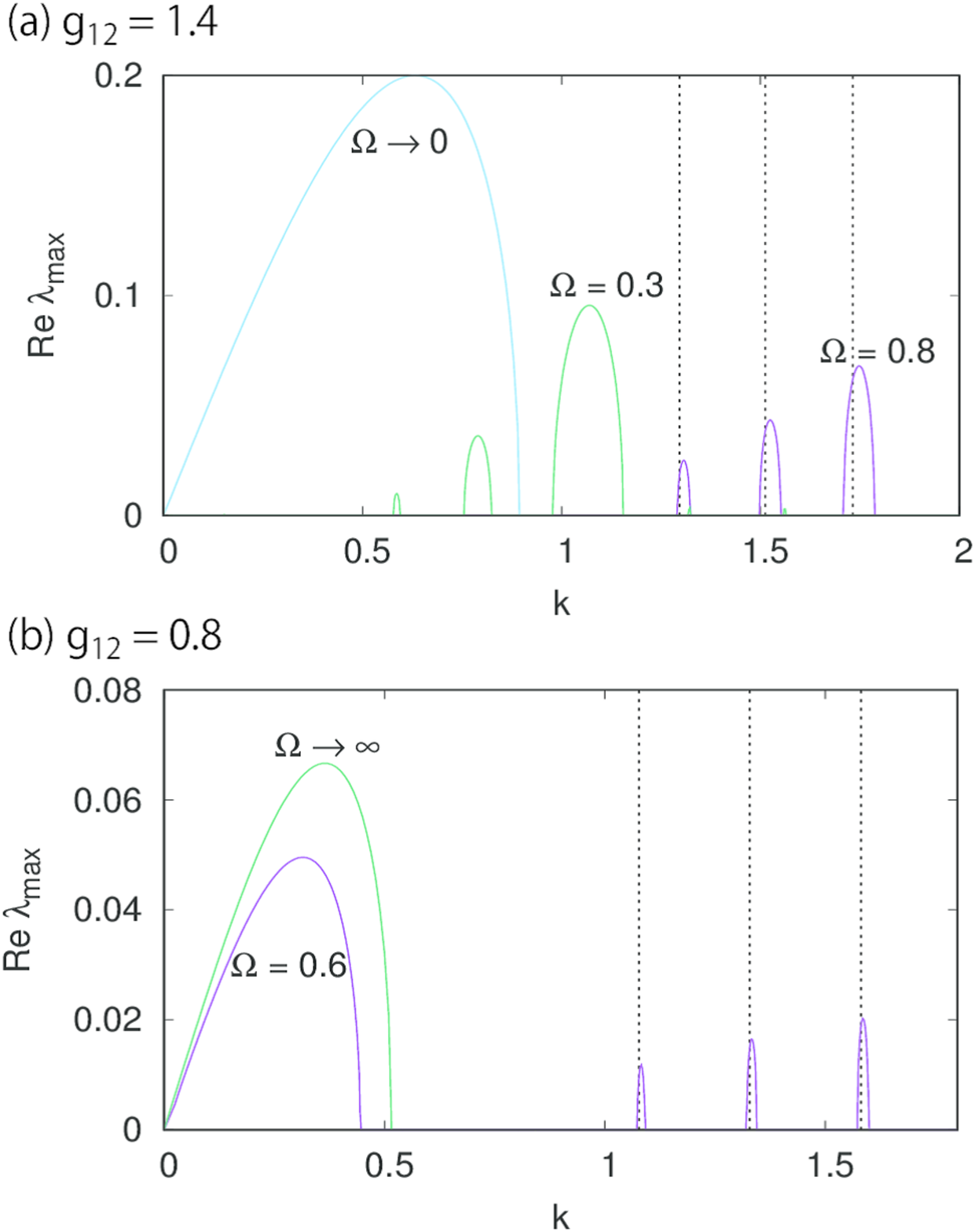}
\caption{
  (color online) Largest real part of the Floquet exponent, ${\rm Re}
  \lambda_{\rm max}$, as a function of wave number $k$, obtained by
  solving Eq.~(\ref{diffuv}) and diagonalizing Eq.~(\ref{UT}) numerically.
  When ${\rm Re} \lambda_{\rm max} > 0$, the system is dynamically unstable
  to excitation of wave number $k$.
  (a) $g_{12} = 1.4$ and (b) $g_{12} = 0.8$.
  In (a) and (b), we plot the imaginary part of $\omega$ in Eqs.~(\ref{om0})
  and (\ref{opposite}) for $\Omega \rightarrow 0$ and $\Omega \rightarrow
  \infty$, respectively.
  The vertical dotted lines are the solutions of Eq.~(\ref{para}).
}
\label{f:floquet}
\end{figure}
We focus on the largest real part of the Floquet exponents $\lambda$ among
the four eigenvalues of $U_T$, which we denote ${\rm Re} \lambda_{\rm max}$.
If ${\rm Re} \lambda_{\rm max}$ is positive, $\bm{u}(t)$ (and therefore
$\delta\psi(\bm{r}, t)$) grows exponentially in time, and the uniform state
is unstable to excitation of wave numbers $k$.
Figure~\ref{f:floquet} shows ${\rm Re} \lambda_{\rm max}$ as a function of
wave number $k$.
In Fig.~\ref{f:floquet}(a), the three peaks for $\Omega = 0.8$ correspond to
the three rings in Fig.~\ref{f:three}.
The three wave numbers in Figs.~\ref{f:three} and \ref{f:floquet}(a) agree
well.
As the Rabi coupling $\Omega$ is decreased, the peaks shift toward smaller
$k$ and the rightmost peak grows while the left two peaks decay, as shown by
the curve for $\Omega = 0.3$ in Fig.~\ref{f:floquet}(a).
In the limit of $\Omega \rightarrow 0$, the left two peaks disappear and the
rightmost peak converges to the imaginary part of $\omega$ in
Eq.~(\ref{om0}).
Thus, for $g_{12} > g$, the instability of Faraday-wave excitation due to
the Rabi oscillation for $\Omega \neq 0$ is continuously connected to the
phase separation for $\Omega = 0$.

In the case of $g_{12} = 0.8$, the uniform state $\psi_1 = \psi_2 = 1 /
\sqrt{2}$ is stable in the absence of the Rabi coupling, and therefore,
${\rm Re} \lambda_{\rm max} = 0$ for $\Omega = 0$. 
In the limit of $\Omega \rightarrow \infty$, on the other hand, the dressed
states become immiscible according to Eq.~(\ref{dressg}).
In Fig.~\ref{f:floquet}(b), we plot the imaginary part of $\omega$ in
Eq.~(\ref{opposite}).
For an intermediate value of $\Omega$, in addition to the dressed-state
instability, the three peaks appear, as shown by the curve of $\Omega = 0.6$
in Fig.~\ref{f:floquet}(b).
Thus, in this parameter regime, the phase-separation instability of the
dressed states coexists with the instability of Faraday-wave excitation
due to the Rabi oscillation.
In the case of $g_{12} = g = 1$, we find that ${\rm Re} \lambda_{\rm max} =
0$ for any $\Omega$, i.e., the uniform state is stable.

\section{Resonance condition}
\label{s:resonance}

In Figs~\ref{f:three}, \ref{f:miscible}, and \ref{f:floquet}, we numerically
found that three wave numbers are excited by the Rabi oscillation.
Here, we clarify the mechanism of this excitation spectrum.

First we consider the case of $g = g_{12}$.
For the initial condition $\psi_1(0) = \psi_2(0) = 1 / \sqrt{2}$, the
uniform solution of the GP equation~(\ref{GP}) is given by
\begin{subequations} \label{gg12uni}
\begin{eqnarray}
\psi_1(t) & = & e^{-i g t} \cos(\Omega t + \pi / 4), \\
\psi_2(t) & = & e^{-i g t} \sin(\Omega t + \pi / 4).
\end{eqnarray}
\end{subequations}
We add a small excitation to this state as
\begin{subequations} \label{dpsi12}
\begin{eqnarray}
\psi_1(\bm{r}, t) & = & e^{-i g t} \left[ \cos(\Omega t + \pi / 4) +
  \delta\psi_1(\bm{r}, t) \right], \\
\psi_2(\bm{r}, t) & = & e^{-i g t} \left[ \sin(\Omega t + \pi / 4) +
  \delta\psi_2(\bm{r}, t) \right].
\end{eqnarray}
\end{subequations}
It is convenient to define $\delta\psi_d(\bm{r}, t)$
and $\delta\psi_s(\bm{r}, t)$ using $\delta\psi_1(\bm{r}, t)$ and
$\delta\psi_2(\bm{r}, t)$ as
\begin{subequations} \label{dpsids}
\begin{eqnarray}
\delta\psi_d & = & \cos(\Omega t + \pi / 4) \delta\psi_1 + \sin(\Omega
t + \pi / 4) \delta\psi_2, \\
\delta\psi_s & = & \cos(\Omega t + \pi / 4) \delta\psi_2 - \sin(\Omega
t + \pi / 4) \delta\psi_1.
\end{eqnarray}
\end{subequations}
Substituting Eqs.~(\ref{dpsi12}) and (\ref{dpsids}) into the GP
equation~(\ref{GP}) and neglecting the second- and third-order terms of
$\delta\psi_d(\bm{r}, t)$ and $\delta\psi_s(\bm{r}, t)$, we obtain the
linearized equations of motion for $\delta\psi_d(\bm{r}, t)$ and
$\delta\psi_s(\bm{r}, t)$ as
\begin{subequations} \label{gamma0}
\begin{eqnarray}
\label{dpsid}
i\frac{\partial \delta\psi_d}{\partial t} & = & -\frac{\nabla^2}{2}
\delta\psi_d + g (\delta\psi_d + \delta\psi_d^*), \\
i\frac{\partial \delta\psi_s}{\partial t} & = & -\frac{\nabla^2}{2}
\delta\psi_s.
\label{dpsis}
\end{eqnarray}
\end{subequations}
Equation~(\ref{dpsid}) has the same form as that of the single-component BEC
and thus gives the well-known Bogoliubov spectrum $\omega =
\sqrt{\epsilon_k (\epsilon_k + 2g)}$.
This excitation modulates the total density, and can be regarded as a
density wave.
Equation~(\ref{dpsis}) gives the free-particle spectrum $\omega =
\epsilon_k$.
Since the interaction coefficient $g$ is not included in
Eq.~(\ref{dpsis}), the excitation occurs with the total density kept
constant, which can be regarded as the spin wave in the quasi-spin picture
of two components.
Since $\Omega$ is not included in both Eqs.~(\ref{dpsid}) and (\ref{dpsis}),
the Rabi oscillation has nothing to do with the excitations for $g =
g_{12}$.

We next consider the case of $g \neq g_{12}$.
We assume that the difference between $g$ and $g_{12}$ is much smaller than
$\Omega$ and take the lowest order of
\begin{equation}
g_{12} - g \equiv \gamma
\end{equation}
in the calculation.
The corrections to the linearized equations of motion (\ref{gamma0}) then
become (see Appendix)
\begin{subequations} \label{short}
\begin{eqnarray}
i\frac{\partial \delta\psi_d}{\partial t} & = & -\frac{\nabla^2}{2}
\delta\psi_d + g (\delta\psi_d + \delta\psi_d^*)
+ \gamma A(\delta\psi_d, \delta\psi_s)
\nonumber \\
& & + \gamma B(\delta\psi_d, \delta\psi_s) \cos 4\Omega t
+ \gamma C(\delta\psi_d, \delta\psi_s) \sin 4\Omega t,
\nonumber \\ \\
i\frac{\partial \delta\psi_s}{\partial t} & = & -\frac{\nabla^2}{2}
\delta\psi_s
+ \gamma A'(\delta\psi_d, \delta\psi_s)
\nonumber \\
& & + \gamma B'(\delta\psi_d, \delta\psi_s) \cos 4\Omega t
+ \gamma C'(\delta\psi_d, \delta\psi_s) \sin 4\Omega t.
\nonumber \\
\end{eqnarray}
\end{subequations}
The functions $A$, $A'$, $B$, $B'$, $C$, and $C'$ in Eq.~(\ref{short})
are linear with respect to $\delta\psi_d$, $\delta\psi_d^*$, $\delta\psi_s$,
and $\delta\psi_s^*$, whose explicit forms are given in Eq.~(\ref{eom}) in
the Appendix.

In the simple parametric resonance~\cite{Landau}, the resonance condition is
determined by two frequencies: the natural frequency of the oscillator
$\omega_0$ and the frequency of the external driving force $\omega_1$.
The parametric resonance occurs when $\omega_1 \simeq 2 \omega_0 = \omega_0
+ \omega_0$, where we neglect the higher-order resonance.
In Eq.~(\ref{short}), the frequency of the driving force $\omega_1$
corresponds to the frequency $4 \Omega$ of the sinusoidal functions, $\sin
4\Omega t$ and $\cos 4\Omega t$.
The natural frequency of the oscillator $\omega_0$ corresponds to the
frequencies that are determined by the first lines in Eq.~(\ref{short}),
i.e., the terms without the sinusoidal functions.
These frequencies $\omega_d$ and $\omega_s$ are obtained in
Eq.~(\ref{omcorr}) in the Appendix.
The frequencies $\omega_d$ and $\omega_s$ in Eq.~(\ref{omcorr}) reduce to
$\omega_d \rightarrow \sqrt{\epsilon_k (\epsilon_k + 2g)}$ and $\omega_s
\rightarrow \epsilon_k$, for $\gamma \rightarrow 0$, and therefore, can be
regarded as the excitation frequencies of the density and spin waves,
respectively.
Since there are two frequencies $\omega_d$ and $\omega_s$, the parametric
resonance can occur for
\begin{subequations} \label{para}
  \begin{eqnarray}
    \label{cond1}
4 \Omega & = & 2 \omega_d, \\
\label{cond2}
4 \Omega & = & 2 \omega_s, \\
\label{cond3}
4 \Omega & = & \omega_d + \omega_s.
\end{eqnarray}
\end{subequations}
For example, the resonance condition in Eq.~(\ref{cond3}) can be understood
as follows.
In the dynamics of Eq.~(\ref{short}), $\delta\psi_d$ and $\delta\psi_s$
acquire a frequency component of $\omega_d$ ($\omega_s$).
This frequency component is multiplied by the sinusoidal functions on the
right-hand side of Eq.~(\ref{short}), which gives a frequency component of
$4\Omega - \omega_d$ ($4\Omega - \omega_s$).
The resonance occurs when this frequency is close to another frequency
component of $\delta\psi_d$ and $\delta\psi_s$, i.e., $\omega_s$
($\omega_d$), which gives Eq.~(\ref{cond3}).

In Figs.~\ref{f:floquet}(a) and \ref{f:floquet}(b), the three vertical
dotted lines show the wave numbers obtained by solving Eq.~(\ref{para}) with
respect to $k$ using Eq.~(\ref{omcorr}).
These analytically obtained wave numbers are in good agreement with the
three numerically obtained peaks.
The leftmost, middle, and rightmost peaks correspond to the resonance
conditions in Eqs.~(\ref{cond1}), (\ref{cond3}), and (\ref{cond2}),
respectively.
It is interesting to note that the rightmost peak is related to the
spin-waves, because it continuously changes into the peak of $\Omega = 0$ in
Fig.~\ref{f:floquet}(a).
This is understood from the fact that the phase separation is triggered by 
the spin-wave excitation for $\Omega = 0$.

We have thus shown that the excitations of multiple wave numbers observed in
Figs~\ref{f:three}, \ref{f:miscible}, and \ref{f:floquet} originate from the
parametric resonance of the density and spin waves with Rabi oscillation.
The three wave numbers come from the combinations of density-density,
spin-spin, and density-spin waves.

\section{Conclusions}
\label{s:conc}

We investigated the dynamics of a two-component BEC with Rabi coupling, and
found that the spatial patterns are excited by the Rabi oscillation.
Due to the Rabi oscillation, the population of each component oscillates,
resulting in the oscillation of the effective interaction.
The oscillation of the interaction causes Faraday-pattern formation through
parametric resonance.
Since there are two kinds of excitations in a two-component BEC, i.e.,
density and spin waves, multiple resonances exist.
We found three resonance peaks in the momentum space (Figs.~\ref{f:three},
\ref{f:miscible}, and \ref{f:floquet}), which originate from the resonance
of the Rabi oscillation with spin-spin, spin-density, and density-density
wave excitations.
We also showed that these parametric resonances coexist with the phase
separation of the dressed states (Figs.~\ref{f:miscible} and
\ref{f:floquet}(b)).

Experimentally, the multiple wave-number excitation presented in this paper
may be observed by time-of-flight expansion with Bragg
spectroscopy~\cite{Vogels}.
The effects of trapping potential and the imbalance in the intracomponent
interaction ($g_{11} \neq g_{22}$) deserve further study.

\begin{acknowledgments}
This work was supported by JSPS KAKENHI Grant Numbers JP17K05595 and
JP17K05596.
\end{acknowledgments}

\appendix

\section{Corrections to the equations of motion for excitations}

In this appendix, we derive Eq.~(\ref{short}), which is the equation of
motion for excitations in the case of $|\gamma| = |g_{12} - g| \ll \Omega$.
We first make a correction to the uniform solution for $g = g_{12}$ in
Eq.~(\ref{gg12uni}) as
\begin{subequations} \label{app1}
\begin{eqnarray}
\psi_1(t) & = & e^{-i (g + \delta\mu) t} \cos(\Omega t + \pi / 4) [1 +
  \Delta_1(t)], \\
\psi_2(t) & = & e^{-i (g + \delta\mu) t} \sin(\Omega t + \pi / 4) [1 +
  \Delta_2(t)],
\end{eqnarray}
\end{subequations}
where $\Delta_1$, $\Delta_2$, and $\delta\mu$ are $O(\gamma)$.
Substituting Eq.~(\ref{app1}) into the GP equation (\ref{GP}) and neglecting
the terms of $O(\gamma^2)$, we find $\delta\mu = \gamma / 4$ and $\Delta_1 =
-\Delta_2 = \gamma / (8\Omega) \cos 2\Omega t$.

We then add small excitations to the uniform wave functions as
\begin{subequations} \label{corr2}
\begin{eqnarray}
  \psi_1(\bm{r}, t) & = & e^{-i(g + \gamma / 4)t} \biggl[
    \cos(\Omega t + \pi / 4)
    \nonumber \\
 & & \times \left( 1 + \frac{i\gamma}{8\Omega} \cos 2\Omega
    t + \delta\psi_1(\bm{r}, t) \right) \biggr], \\
  \psi_2(\bm{r}, t) & = & e^{-i(g + \gamma / 4)t} \biggl[
    \sin(\Omega t + \pi / 4)
    \nonumber \\
& & \times \left( 1 - \frac{i\gamma}{8\Omega} \cos 2\Omega
    t + \delta\psi_2(\bm{r}, t) \right) \biggr].
\end{eqnarray}
\end{subequations}
We substitute these wave functions into the GP equation~(\ref{GP}) and
neglect the terms of $O(\gamma^2)$.
After a straightforward calculation, we obtain
\begin{subequations} \label{eom}
\begin{eqnarray}
\label{eom1}
i\frac{\partial \delta\psi_d}{\partial t} & = & -\frac{\nabla^2}{2}
\delta\psi_d + g (\delta\psi_d + \delta\psi_d^*)
\nonumber \\
& & + \frac{\gamma}{16} \biggl\{
  4 (\delta\psi_d + \delta\psi_d^*) + \frac{ig}{\Omega} (\delta\psi_s -
  \delta\psi_s^*)
\nonumber \\
& &  + \left[ 4(2\delta\psi_d + \delta\psi_d^*) + \frac{ig}{\Omega}
    (\delta\psi_s - \delta\psi_s^*) \right] \cos 4\Omega t
\nonumber \\
& & - \left[ \frac{2ig}{\Omega} \delta\psi_d^* + 4(2\delta\psi_s +
  \delta\psi_s^*) \right] \sin 4\Omega t \biggr\}, \\
i\frac{\partial \delta\psi_s}{\partial t} & = & -\frac{\nabla^2}{2}
\delta\psi_s
\nonumber \\
& & + \frac{\gamma}{16} \biggl\{
-\frac{ig}{\Omega} (\delta\psi_d + \delta\psi_d^*)
 + 4 (\delta\psi_s - \delta\psi_s^*)
\nonumber \\
& &  - \left[ \frac{ig}{\Omega} (\delta\psi_d + \delta\psi_d^*)
+ 4(2\delta\psi_s + \delta\psi_s^*) \right] \cos 4\Omega t
\nonumber \\
& & - 4(2\delta\psi_d + \delta\psi_d^*) \sin 4\Omega t \biggr\},
\label{eom2}
\end{eqnarray}
\end{subequations}
These equations of motion reduce to Eq.~(\ref{gamma0}) for $\gamma = 0$.

If we take the first and second lines in Eqs.~(\ref{eom1}) and (\ref{eom2}),
i.e., if we drop the terms with sinusoidal functions, the Bogoliubov matrix
becomes
\begin{equation}
  \left( \begin{array}{cccc}
    \epsilon_k + g + \frac{\gamma}{4} & g + \frac{\gamma}{4} &
    \frac{ig\gamma}{16\Omega} & \frac{ig\gamma}{16\Omega} \\
    -g - \frac{\gamma}{4} & -\epsilon_k - g - \frac{\gamma}{4} &
    \frac{ig\gamma}{16\Omega} & \frac{ig\gamma}{16\Omega} \\
    -\frac{ig\gamma}{16\Omega} & -\frac{ig\gamma}{16\Omega} &
    \epsilon_k + \frac{\gamma}{4} & -\frac{\gamma}{4} \\
    -\frac{ig\gamma}{16\Omega} & -\frac{ig\gamma}{16\Omega} &
    \frac{\gamma}{4} & -\epsilon_k - \frac{\gamma}{4} \end{array} \right).
\end{equation}
The eigenvalues of this matrix give the excitation spectrum, and are
obtained as
\begin{subequations} \label{omcorr}
\begin{eqnarray}
\omega_d & = & \sqrt{\epsilon_k (\epsilon_k + 2g)} + \frac{\epsilon_k
  \gamma}{4\sqrt{\epsilon_k (\epsilon_k + 2g)}} + O(\gamma^2), \\
\omega_s & = & \epsilon_k + \frac{\gamma}{4} + O(\gamma^2).
\end{eqnarray}
\end{subequations}
These excitation frequencies reduce to $\omega_d = \sqrt{\epsilon_k
  (\epsilon_k + 2g)}$ and $\omega_s = \epsilon_k$ for $\gamma = 0$.

\end{document}